\documentstyle[11pt,newpasp,twoside]{article}
\def\edcomment#1{\iffalse\marginpar{\raggedright\sl#1\/}\else\relax\fi}
\reversemarginpar

\begin{document}
\title{Galaxy Scaling Relations as a Result of
Secular Evolution}
\author{Xiaolei Zhang}
\affil{US Naval Research Laboratory, 4555 Overlook Ave SW,
Washington, DC 20375, USA}

\begin{abstract}
The secular evolution process which slowly transforms
the morphology of a given galaxy over its lifetime
through mostly internal dynamical mechanisms
could naturally account for
most of the observed properties of physical galaxies (Zhang 2003a).
As an emerging paradigm for galaxy evolution, its dynamical
foundations had been established in the past few years 
(Zhang 1996, 1998, 1999).
In this paper, we explore further implications of the secular
morphological evolution process in reproducing the well-known
scaling relations of galaxies.
\end{abstract}

\section{Disk Galaxy Scaling Relations}

A typical galaxy remains in a quasi-equilibrium configuration
during the secular evolution process.
From the Virial theorem relation $V^2 = GM_{dyn}/R$, 
where $M_{dyn}$ is the dynamical mass of the galaxy.
and the definition of average surface brightness $SB=L/R^2$, where
L is the luminosity, it follows that
$ L \propto V^4 {{1} \over {SB}} {{1} \over {(M_{dyn}/L)^2}},$
In order to have a tight Tully-Fisher relation $L \propto V^4$
(Tully \& Fisher 1977),
we must have $SB (M_{dyn}/L)^2 \approx constant$.
The secular evolution process maintains the scaling relations
by a decrease in galaxy's dynamical-mass-to-light ratio
as the surface brightness of the galaxy increases during
its Hubble type transformation (Zhang 2003b).

The fundamental plane relation for spirals
can likewise be derived from the Virial theorem,
resulting in $ 10 \log V = - (1+2\beta) M_t - SB + constant,
$ where $M_t$ is the absolute magnitude
and $SB$ is the average surface brightness in magnitude/arcsec$^2$,
and where $M_{dyn}$/L $\propto L^{\beta}$.
Pharasyn et al. (1997) found that fitting I band and K band data
of a sample of spiral galaxies to this relation resulted in
$\beta \approx -0.15$. 

\section{Elliptical Galaxy Fundamental Plane Relations}

One of the many forms of the elliptical galaxy
fundamental plane relation
can be expressed as $ L \sim V_e^{3.45} SB^{-0.86},$ or
$ M(R_e) = -8.62 (\log V + 0.1 SB) + constant$ (Djorgovski \& Davis 1987),
where $M(R_e)$ refers to the absolute magnitude inside effective radius 
$R_e$.  A slight rearrangement of the terms in this
equation leads to $ 10 \log V = -1.25 M(R_e) - SB + constant.$
We see that apart from the constant term, the only difference
between the spiral and elliptical fundamental plane
relations is in the different signs of the mass-to-light
ratio exponent $\beta$: $\beta=-0.15$ for spirals
and $\beta = 0.13$ for ellipticals and bulges.
As the secular evolution proceeds (which generally leads to
an increase in L),
the dynamical mass-to-light ratio of a spiral galaxy decreases
because a larger fraction of the baryonic dark matter
becomes luminous.  On the other hand, elliptical galaxies
are in more advanced stages of evolution
and will experience greater degree of dimming,
which is reflected in the increase of dynamical mass-to-light
ratio with L.

\section{The Evolution Trend Revealed by the
Three Types of Scaling Relations}

From the above three types of scaling relations, 
we can easily derive the
following variations of the dependence of M$_{dyn}$/L on mass
and on light, and the inter-dependence of M$_{dyn}$ and L:

(1). The dependence of mass on luminosity:
$M_{dyn} \sim L^{0.5}$ from spiral Tully-Fisher relation,
$M_{dyn} \sim L^{0.85}$ from spiral fundamental plane relation,
$M_{dyn} \sim L^{1.12}$ from elliptical fundamental plane relation.

(2). The dependence of M$_{dyn}$/L on luminosity:
$M_{dyn}/L \sim L^{-0.5}$ from spiral Tully-Fisher relation,
$M_{dyn}/L \sim L^{-0.15}$ from spiral fundamental plane relation,
$M_{dyn}/L \sim L^{0.13}$ from elliptical fundamental plane relation.

(3). The dependence of M/L on dynamical mass:
$M_{dyn}/L \sim M_{dyn}^{-1}$ from spiral Tully-Fisher relation,
$M_{dyn}/L \sim M_{dyn}^{-0.18}$ from spiral fundamental plane relation,
$M_{dyn}/L \sim M_{dyn}^{0.1}$ from elliptical fundamental plane relation.

A gradual evolution trend of galaxy properties is thus implied
in the three types of scaling relations.
The fact that the M$_{dyn}$/L ratio obtained
from the spiral galaxy fundamental plane relation
lies intermediate between that given by the spiral galaxy
Tully-Fisher relation and the elliptical
galaxy fundamental plane relation is apparently due to the fact that
within the spiral fundamental plane relation there is
the additional surface brightness parameter.
The spiral fundamental
plane effectively samples disk galaxy properties further
inward, compared to the
Tully-Fisher relation which samples characteristics
averaged over the whole disk.
More detailed discussion of the relation of the secular evolution
process to the generation of galaxy structural properties and 
scaling relations can be found in Zhang (2003b).

\end{document}